\documentclass[superscriptaddress,aps,showpacs,nofootinbib,twocolumn]{revtex4}
\usepackage{graphicx}
\usepackage{amsmath,amssymb}

\def\GRG{{\it Gen. Relativity and Gravitation} }

\def\PL{{\it Phys. Lett.} }
\def\PR{{\it Phys. Rev.} }
\def\PRL{{\it Phys. Rev. Lett.} }

\def\PTP{{\it Progr. Theor. Phys.} }

\def\vev#1{\langle {#1}\rangle}
\def\bra#1{\langle{#1}|}
\def\ket#1{|{#1}\rangle}

\def\frac#1#2{{\textstyle{{#1}\over {#2}}}}

\def\lsim{\mathrel{\rlap{\lower4pt\hbox{\hskip1pt$\sim$}}
    \raise1pt\hbox{$<$}}}
\def\gsim{\mathrel{\rlap{\lower4pt\hbox{\hskip1pt$\sim$}}
    \raise1pt\hbox{$>$}}}
\def\sqr#1#2{{\vcenter{\vbox{\hrule height.#2pt
         \hbox{\vrule width.#2pt height#1pt \kern#1pt
         \vrule width.#2pt}
         \hrule height.#2pt}}}}

\def\beq{\begin{equation}}
\def\eeq{\end{equation}}
\def\beqa{\begin{eqnarray}} 
\def\eeqa{\end{eqnarray}}

\def\laq{\raise 0.4 ex \hbox{$<$}\kern -0.8 em\lower 0.62 ex\hbox{$\sim$}}
\def\gaq{\raise 0.4 ex \hbox{$>$}\kern -0.7 em\lower 0.62 ex\hbox{$\sim$}}

\begin{document}

\title{Bounds on Cubic Lorentz-Violating Terms in the Fermionic Dispersion Relation}

\author{O. Bertolami}

\altaffiliation {Email address: orfeu@cosmos.ist.utl.pt }

\author{J. G. Rosa}

\altaffiliation[] {Email address: joaopedrotgr@sapo.pt}
 
\affiliation{ Departamento de F\'\i sica, Instituto Superior T\'ecnico \\
Avenida Rovisco Pais 1, 1049-001 Lisboa, Portugal}

%\vskip 0.5cm

\date{\today}

\begin{abstract}

We study the recently proposed Lorentz-violating dispersion relation for fermions and 
show that it leads to two distinct cubic operators in the momentum. We compute the 
leading order terms that modify the non-relativistic equations of motion and 
use experimental results for the hyperfine transition in the ground state of the ${}^9\textrm 
Be^+$ ion to bound the values of the Lorentz-violating parameters $\eta_1$ and $\eta_2$ for 
neutrons. The resulting bounds depend on the value of the 
Lorenz-violating background four-vector in the laboratory frame.

\end{abstract}

\pacs{ 11.30.Cp, 04.60.-m, 11.10.Ef \hspace{2cm}Preprint DF/IST-9.2004}

\maketitle
%\vskip 2pc
 
%%%%%%%%%%%%%%%%%%%%%%%%%%%%%%%%%%%%

\section{Introduction}

The possibility of violation of the Lorentz symmetry has been widely discussed in the recent literature (see 
e.g. \cite{Kostelecky1}). Indeed, the spontaneous breaking of this fundamental symmetry 
may arise in the context of string/M-theory due to existence of
non-trivial solutions in string field theory  \cite{Kostelecky2}, 
in loop quantum gravity \cite{Gambini,Alfaro}, 
in noncommutative field theories 
\cite{Carroll} \footnote{Notice however, that in a model where a scalar field is coupled to gravity, Lorentz 
invariance may hold, at least at first non-trivial order in perturbation theory of the 
noncommutative parameter \cite{Bertolami}.}, in quantum gravity inspired spacetime foam 
scenarios \cite{Garay} or through the spacetime variation of fundamental coupling constants \cite{Lehnert}. 
This breaking could be tested, for instance, in ultra-high energy cosmic rays \cite{Sato}. 

Recently, it has been proposed a method of introducing cubic modifications into dispersion relations 
by means of dimension five operators for fermions \cite{Myers}. The upper bounds for the parameters 
that characterize these modifications are based on low-energy experiments, being $|\xi|\lesssim10^{-6}$ 
for the electromagnetic sector, $|\eta_{Q,u,d}|\lesssim10^{-6}$ for first quark generation 
and $|\eta^e_{L,R}|\lesssim10^{-5}$ for electrons \cite{Myers}.

In this Letter, we shall consider cubic Lorentz-violating terms for fermions in the non-relativistic 
limit and obtain new upper bounds for neutrons, based on spectroscopical results for the ${}^9\textrm Be^+$ 
ground state, as discussed by Bollinger \emph{et al.} \cite{Bollinger}.

%%%%%%%%%%%%%%%%%%%%%%%%%%%%%%%%%%%%%%%%%%%%%%%%%%%%%%%%%%%%%%%%%%

\section{The model}

We consider terms in the Lagrangian density which describes a Dirac spinor field, 
correspondig to dimension five operators which break the Lorentz symmetry by means 
of a background four-vector $n^{\mu}$ \cite{Myers}. These terms have the following features: 
(i) have one more derivative than the usual kynetic term, 
(ii) are gauge invariant, (iii) are Lorentz invariant, 
apart from  $n^{\mu}$, (iv) are irreducible to lower dimension operators by means of the equations of 
motion and (v) do not correspond to a total derivative and are suppressed by a single power of the Planck mass, 
$M_P$.

Under these conditions, the two possible operators can be combined in the following form \cite{Myers}:
\begin{equation} \label{dirac_lagrangian_1}
\mathcal{L}_f={1\over M_P}\bar{\psi}(\eta_1\not\!n+\eta_2\not\!n\gamma_5)(n\cdot\partial)^2\psi~~.
\end{equation}

The parameters $\eta_1$ and $\eta_2$ can, for instance, in the case of string theory, 
be regarded as vacuum expectation values of tensor operators 
arising from the spontaneous symmetry breaking mechanism \cite{Kostelecky2}.

First, it should be pointed out that the Lagrangian density Eq. (\ref{dirac_lagrangian_1}) is not 
symmetric in what respects the fields $\psi$ and $\bar{\psi}$ and, thus, one should include its hermitian conjugate. 
The complete fermionic Lagrangian density is, hence, given by
\begin{eqnarray} \label{dirac_lagrangian_2}
\mathcal{L}_f&=&\bar{\psi}(i\not\!\partial-m)\psi+\qquad\qquad\nonumber\\
&\:&+{1\over M_P}\bar{\psi}(\eta_1\not\!n+\eta_2\not\!n\gamma_5)(n\cdot\partial)^2\psi+\textrm{h.c.}~~,
\end{eqnarray}
which must satisfy the following Euler-Lagrange equations:
\begin{equation} \label{lagrange_equation}
{\partial\mathcal{L}\over\partial\varphi}-\partial_{\mu}\bigg({\partial\mathcal{L}\over\partial(\partial_{\mu}\varphi)}\bigg)
+\partial_{\mu}\partial_{\nu}\bigg({\partial\mathcal{L}\over\partial(\partial_{\mu}\partial_{\nu}\varphi)}\bigg)=0~~,
\end{equation}
where $\varphi$ denotes a generic field of the Lagrangian density. For $\varphi=\bar{\psi}$, 
Eq. (\ref{lagrange_equation}) leads to the modified Dirac equation:
\begin{equation} \label{dirac_equation_1}
\left[i\not\!\partial-m+{1\over M_P}(\eta_1\not\!n+\eta_2\not\!n\gamma_5)(n\cdot\partial)^2\right]\psi=0~~.
\end{equation}
For $\varphi=\psi$, we obtain, as expected, the hermitian conjugated equation. 

In order to obtain the correspondent dispersion relation, we operate Eq. (\ref{dirac_equation_1}) 
with ($i\not\!\partial+m+\frac{1}{M_P}(\eta_1\not\!n+\\\eta_2\not\!n\gamma_5)(n\cdot\partial)^2$), and after
neglecting terms of order $M_P^{-2}$ we obtain, by using
\begin{eqnarray}\label{commutation_relations}
\{\not\!\partial,\not\!n\}&=&2(n\cdot\partial),\\
\{\not\!\partial,\not\!n\gamma_5\}&=&[\not\!\partial,\not\!n]\gamma_5=-2i\gamma_5\sigma^{\mu\nu}n_{\nu}\partial_{\mu}~~,
\end{eqnarray}
that
\begin{equation} \label{dispersion_relation_2}
\big(\Box+m^2\big)\psi=
{2i\over M_P}\big(\eta_1(n\cdot\partial)^3-i\eta_2\gamma_5\sigma^{\mu\nu}n_{\nu}\partial_{\mu}(n\cdot\partial)^2\big)\psi~.
\end{equation}
\noindent
Finally, in the frame where $n^{\mu}=(1,0,0,0)$, we find the dispersion relation
\begin{equation} \label{dispersion_relation_3}
E^2-|\vec{p}|^2-m^2-{2\over M_P}(\eta_1E^3+i\eta_2\gamma_5\sigma^{0\mu}p_{\mu}E^2)=0~~.
\end{equation}
\noindent
Thus, we conclude that the terms in $\eta_1$ and $\eta_2$ yield two different cubic modifications in the 
momentum operator of the fermionic dispersion relation. The first one is similar to the one of Ref. \cite{Myers}, while 
the second is a new term identified here for the first time. 

%%%%%%%%%%%%%%%%%%%%%%%%%%%%%%%%%%%%%%%%%%%%%%%%%%%%%%%%%%%%%%%%%%%%%%%%%%%%%%%%%%%%%%%%%%%%%%%%%%%%%%

\section{The non-relativistic limit and the ${}^9\textrm{Be}^+$ ion energy spectrum}

Let us now determine how the Lorentz-violating terms in Eq. (\ref{dirac_equation_1}) affect 
the equations of motion in the non-relativistic limit. For this, we can write the four component spinor $\psi$ in the form
\begin{equation} \label{spinor_1}
\psi=\left(\begin{array}{ccc}
\hat{\varphi}\\
\hat{\chi}
\end{array}\right)~~,
\end{equation}
where $\hat{\varphi}$ and $\hat{\chi}$ are two component spinors. 
Eq. (\ref{dirac_equation_1}) can be, thus, written as a system of two equations:
\begin{eqnarray}\label{non_rel_dirac_1a}
i\partial_0\hat{\varphi}+i(\vec{\sigma}\cdot\vec{\nabla})\hat{\chi}-m\hat{\varphi}&=&
-{1\over M_P}\big[A(n\cdot\partial)^2\hat{\varphi}+\nonumber\\
&\:&+B(n\cdot\partial)^2\hat{\chi}\big]~~,
\end{eqnarray}
\begin{eqnarray}\label{non_rel_dirac_1b}
i\partial_0\hat{\chi}+i(\vec{\sigma}\cdot\vec{\nabla})\hat{\varphi}
+m\hat{\chi}&=&-{1\over M_P}\big[A(n\cdot\partial)^2\hat{\chi}+\nonumber\\
&\:&+B(n\cdot\partial)^2\hat{\varphi}\big]~~,
\end{eqnarray}
where $A\equiv\eta_1n_0-\eta_2(\vec{n}\cdot\vec{\sigma})$ and $B\equiv\eta_2n_0-\eta_1(\vec{n}\cdot\vec{\sigma})$. 
In the low-energy limit, $E-m \ll m$, and we can separate the slowly and the rapidly time-varying parts of spinors 
$\hat{\varphi}$ and $\hat{\chi}$ in the following way:
\begin{equation} \label{spinor_2}
\left(\begin{array}{ccc}
\hat{\varphi}\\
\hat{\chi}
\end{array}\right)=e^{-imt}\left(\begin{array}{ccc}
\varphi\\
\chi
\end{array}\right)~~.
\end{equation}
Hence, Eqs. (\ref{non_rel_dirac_1a}) and (\ref{non_rel_dirac_1b}) become:
\begin{equation}\label{non_rel_dirac_2a}
i\partial_0\varphi+i(\vec{\sigma}\cdot\vec{\nabla})\chi=-{1\over M_P}\big[A(F\varphi)+B(F\chi)\big]~~,
\end{equation}
\begin{equation}\label{non_rel_dirac_2b}
i\partial_0\chi+i(\vec{\sigma}\cdot\vec{\nabla})\varphi+2m\chi=-{1\over M_P}\big[A(F\chi)+B(F\varphi)\big]~~,
\end{equation}
where the operator $F$ is given by
\begin{eqnarray} \label{operator_F}
F&=&{n_0}^2(\partial_0^2-2im\partial_0-m^2)+\nonumber\\
&\:&+2n_0(-im+\partial_0)(\vec{n}\cdot\vec{\nabla})+(\vec{n}\cdot\vec{\nabla})^2~~.
\end{eqnarray}

As we are looking for the leading order terms for Lorentz violation in the non-relativistic 
limit, we can neglect terms of order $M_P^{-1}$ in Eq. (\ref{non_rel_dirac_2b}) in order to 
obtain a zeroth-order relation between the spinors $\varphi$ and $\chi$. As $\chi$ varies slowly 
in time, we can also neglect its time derivative, and so
\begin{equation} \label{spinors_relation}
\chi\approx{-i(\vec{\sigma}\cdot\vec{\nabla})\over 2m}\varphi={(\vec{\sigma}\cdot\vec{p})\over 2m}\varphi\ll\varphi~~.
\end{equation}
Substituting this result into Eq. (\ref{non_rel_dirac_2a}) and neglecting terms of order $m/M_P$ 
and $m^2/M_P$, as well as those terms which include time derivatives of the spinors that are 
supressed by the Planck mass $M_P$, we obtain
\begin{eqnarray} \label{modified_schrodinger}
i\partial_0\varphi&=&{1\over 2m}\nabla^2\varphi-{1\over M_P}\bigg[A(\vec{n}\cdot\vec{\nabla})^2-\nonumber\\
&\:&-{i\over 2m}B(\vec{n}\cdot\vec{\nabla})^2(\vec{\sigma}\cdot\vec{\nabla})\bigg]\varphi~~.
\end{eqnarray}

We have then found the two leading order terms that modify the kynetic term of the 
Schr\"odinger equation for the positive energy spinor $\varphi$. In general, these terms will 
modify the Hamiltonian for a system of $N$ particles through a Lorentz-violating potential given by:
\begin{eqnarray} \label{perturbative_potential}
\hat{V}&=&-{1\over M_P}\sum_{k=1}^{N}\bigg[\big(\eta_1n_0-\eta_2(\vec{n}\cdot\vec{\sigma})\big)(\vec{n}\cdot\vec{\nabla}_k)^2-\nonumber\\
&\:&-{i\over2m_k}\big(\eta_2n_0-\eta_1(\vec{n}\cdot\vec{\sigma})\big)(\vec{n}\cdot\vec{\nabla}_k)^2(\vec{\sigma}\cdot\vec{\nabla}_k)\bigg]
\end{eqnarray}
where $\vec{\nabla}_k\equiv\partial/\partial\vec{r}_k$, and $\vec{r}_k$, $k=1,\ldots,N$, 
is the position vector of the \emph{k-th} particle with mass $m_k$, respectively.

In a 1989 paper, Steven Weinberg proposed the use of a hyperfine transition in the ground 
state of the ${}^9\textrm Be^+$ ion to test a non-linear generalization of quantum mechanics 
\cite{Weinberg}. Although we are looking for the effects of linear Lorentz-violating operators in the 
Schr\"odinger equation, Weinberg's method can be easily adapted to our purposes.

Consider a system in a coherent superposition of two quantum states, $\psi_1$ and $\psi_2$, 
whose energy eigenvalues in the absence of Lorentz violation are $E_1$ and $E_2$, respectively. 
This system is described by the Hamiltonian $\hat{H}=\hat{H}_0+\hat{V}$, where $\hat{V}$ can be 
treated as a perturbative potential compared to the system's Lorentz invariant Hamiltonian 
$\hat{H}_0$, as we expect the effects of the Lorentz invariance violation to be small at this 
energy scale. To first order in perturbation theory, the Schr\"odinger time-dependent equation 
for state $\psi_k$, $k=1,2$, takes the form
\begin{equation} \label{perturbed_schrodinger}
i\hbar{\partial \psi_k\over\partial t}=(E_k+\vev{\hat{V}}_k)\psi_k=\hbar\omega_k\psi_k ~~,
\end{equation}
where $\vev{\hat{V}}_k\equiv\bra{\psi_k}\hat{V}\ket{\psi_k}$, and has the general solution $\psi_k=c_ke^{-i\omega_kt}$.

The constants $c_k$ can be parametrized as $c_1=\sin(\frac{\theta}{2})$ and 
$c_2=\cos(\frac{\theta}{2})$ \cite{Bollinger}. The relative phase of the two states, 
correspondent to the time dependence of $\psi_2^{\dag}\psi_1$, is given by
\begin{equation} \label{precession_frequency_1}
\omega_p\equiv\omega_1-\omega_2=\omega_0+{\vev{\hat{V}}_1-\vev{\hat{V}}_2\over\hbar}~~,
\end{equation}
where $\omega_0\equiv(E_1-E_2)/\hbar$ is the frequency of the transition between the unperturbed states. 
The perturbative terms will, thus, depend on the parameter $\theta$ and, hence, measuring the $\theta$ 
dependence of $\omega_p$ allows for determining the effects of the Lorentz invariance violation on the system.

A two level system is mathematically equivalent to a spin $1/2$ system which undergoes 
precession about an external uniform magnetic field, with $\theta$ being the angle between 
the spin and magnetic field vectors and $\omega_p$ the precession frequency. Bollinger \emph{et al.} 
have used this idea to search for a $\theta$ dependence of the precession frequency of the hyperfine 
transition $\ket{m_I,m_J}=\ket{-\frac{1}{2},\frac{1}{2}}\rightarrow\ket{-\frac{3}{2},\frac{1}{2}}$ 
in the ground state of the ${}^9\textrm Be^+$ ion \cite{Bollinger}.

In their discussion it has been assumed that the ${}^9\textrm Be^+$ nuclear spin was 
decoupled from the valence electron's spin, so that $\psi_1\equiv\ket{-\frac{3}{2},\frac{1}{2}}$ 
and $\psi_2\equiv\ket{-\frac{1}{2},\frac{1}{2}}$ are pure $\ket{m_I,m_J}$ states. 
With this hypothesis, they obtained the upper bound
\begin{equation} \label{bollinger_result}
\bigg|{\omega_p(\theta_B)-\omega_p(\theta_A)\over 2\pi}\bigg|\leq  12.1\ \mathrm{\mu Hz}
\end{equation}
for $\theta_A=1.02$ rad and $\theta_B=2.12$ rad.

To determine how the breaking of the Lorentz symmetry produces a 
$\theta$ dependence in $\omega_p$, we have to compute the expectation value 
of the perturbative potential on states $\psi_1$ and $\psi_2$. We first point out that
\begin{equation} \label{B_term}
(\vec{n}\cdot\vec{\nabla})^2(\vec{\sigma}\cdot\vec{\nabla})=-in^in^j\sigma^kp^ip^jp^k~~,
\end{equation}
where $p^i$ is the \emph{i-th} component of the vector momentum. As, for bound states like 
$\psi_1$ and $\psi_2$, any odd power of the momentum operator has a zero expectation value, 
the term in B will not affect the perturbative potential's expectation value \cite{Kostelecky_2}.

The ${}^9\textrm Be^+$ ion is a system composed by three electrons, two of which in a closed 
$1s$ shell, and a nucleus with five neutrons and four protons. As, in the considered transition, 
$\Delta m_J=0$, we expect the perturbative potential to alter both states energy eigenvalues in 
the same way, not affecting the transition frequency. In the ion's nucleus, the 
\emph{pairing interaction} induces nucleons to group up into pairs of neutrons and 
pairs of protons with zero angular momentum \cite{Wong}. Hence, the ion's nuclear spin is 
entirely carried by one of its neutrons.

In this way, $\psi_1$ and $\psi_2$ can be treated as states of a particle with spin 
$I=3/2$ and projections on the quantization axis, which is usually defined as the 
external magnetic field's direction, $m_I=-3/2$ and $m_I=-1/2$, respectively. 
If $\hat{e}_3$ defines the direction of the quantization axis, 
\begin{equation} \label{spin_expected_value}
\bra{I,m_I}\sigma^k\ket{I,m_I}=2m_I\delta_{k3}~~,
\end{equation}
and therefore
\begin{eqnarray}\label{potential_expected_values}
\vev{\hat{V}}_1&=&{|c_1|^2\over M_P}[\eta_1n_0+3\eta_2n_z]n^in^j\vev{p^ip^j}_1~~,\\
\vev{\hat{V}}_2&=&{|c_2|^2\over M_P}[\eta_1n_0+\eta_2n_z]n^in^j\vev{p^ip^j}_2~~.
\end{eqnarray}
\noindent
Hence, we find (inserting back the missing $h$ factors)
\begin{eqnarray} \label{precession_frequency_2}
\omega_p(\theta)&=&\omega_0-{n^in^j\vev{p^ip^j}\over M_P\hbar}\big[\eta_1n_0\big(\cos^2(\theta/2)-\nonumber\\
&\:&-\sin^2(\theta/2)\big)+\eta_2n_z\big(\cos^2(\theta/2)-\nonumber\\
&\:&-3\sin^2(\theta/2)\big)\big]~~,
\end{eqnarray}
where we have assumed that $\vev{p^ip^j}\equiv\vev{p^ip^j}_1\approx\vev{p^ip^j}_2$. Finally, we obtain
\begin{equation} \label{precession_frequency_3}
{\omega_p(\theta_B)-\omega_p(\theta_A)\over2\pi}={n^in^j\vev{p^ip^j}\over hM_P}\big[a\eta_1n_0+b\eta_2n_z\big] ~~,
\end{equation}
where the constants $a$ and $b$ are defined as
\begin{eqnarray} \label{constants}
a&\equiv&\cos(\theta_A)-\cos(\theta_B)\simeq1.045~~,\\
b&\equiv&-\cos^2\big(\frac{\theta_B}{2}\big)+3\sin^2\big(\frac{\theta_B}{2}\big)+\nonumber\\
&\:&+\cos^2\big(\frac{\theta_A}{2}\big)-3\sin^2\big(\frac{\theta_A}{2}\big)\simeq 2.091~~.
\end{eqnarray}

As for a neutron, $\vev{p^2}/m_n^2\sim10^{-2}$ \cite{Kostelecky_2}, 
and assuming that the Lorentz symmetry breaking does not privilege any spatial direction, $n_x=n_y=n_z\equiv n$, we obtain:
\begin{equation} \label{momentum_expected_value}
{n^in^j\vev{p^ip^j}\over hM_P}\sim{9n^2\vev{p^2}\over hM_P}\sim(2\times10^3)n^2\ \mathrm{Hz}~~.
\end{equation}
%%%%%%%%%%%%%%%%%%%%%%%%%%%%%%%%%%%%%%%%%%%%%%%%%%%%%%%%%%%%%%%%%%%%%%%%%%%%%%%%%%%%%%%%%%%%%%%%%%%%

\section{Results}

As presently there is no way of determining the form of the background four-vector 
$n^{\mu}$, we can only estimate bounds on the values of the parameters $\eta_1$ and $\eta_2$. 

First, we consider the case where $n^{\mu}$ is a time-like four-vector in some cosmic frame 
($n\cdot n=1$). Thus, in the laboratory frame, $n_0\sim 1$ and the typical size of the spatial 
components will be of order $n\sim 10^{-3}$ to the relative motion of our galaxy, 
the Solar System and the Earth \cite{Myers, Mocioiu}. Hence,
\begin{equation} \label{precession_frequency_4}
{\omega_p(\theta_B)-\omega_p(\theta_A)\over 2\pi}\simeq(2\times 10^{-3}\eta_1+4.5\times 10^{-6}\eta_2)\ \mathrm{Hz}~.
\end{equation}

Using Bollinger \emph{et al.} result Eq. (\ref{bollinger_result}), 
we obtain the following upper bounds for the Lorentz-violating parameters:
\begin{equation} \label{parameters_time_like}
|\eta_1|\lesssim6\times10^{-3}\qquad,\qquad |\eta_2|\lesssim3~~,
\end{equation}
where we have assumed $\eta_1(\eta_2)=0$ to obtain a bound for $\eta_2(\eta_1)$.

If $n^{\mu}$ is space-like in some cosmic frame ($n\cdot n$= -1), we will have, 
in the laboratory frame, $n_0\sim10^{-3}$ and $n\sim\sqrt{3}/3$. Thus,
\begin{equation} \label{precession_frequency_5}
{\omega_p(\theta_B)-\omega_p(\theta_A)\over 2\pi}\simeq(0.76\eta_1+8.7\times10^2\eta_2)\ \mathrm{Hz}~,
\end{equation}
and, in this case, we obtain the upper bounds
\begin{equation} \label{parameters_space_like}
|\eta_1|\lesssim2\times10^{-5}\qquad,\qquad |\eta_2|\lesssim1\times10^{-8}~~.
\end{equation}
Finally, considering the case where $n^{\mu}$ is a light-like four-vector in the laboratory 
frame ($n\cdot n=0$), with $n_0\sim1$ and $n\sim\sqrt{3}/3$, we get
\begin{equation} \label{precession_frequency_6}
{\omega_p(\theta_B)-\omega_p(\theta_A)\over2\pi}\sim(7.5\times10^2\eta_1+8.7\times10^2\eta_2)\ \mathrm{Hz}~,
\end{equation}
and the correspondent upper bounds
\begin{equation} \label{parameters_light_like}
|\eta_1|\lesssim2\times10^{-8}\qquad,\qquad |\eta_2|\lesssim1\times10^{-8} ~~.
\end{equation}

%\vskip 0.2cm
%%%%%%%%%%%%%%%%%%%%%%%%%%%%%%%%%%%%%%%%%%%%%%%%%%%%%%%%%%%%%%%%%%%%%%%%%%%%%%%%%%%%%%%%%%%%%%%%%%%%%%%%%%%%

\section{Conclusions}

In this Letter, we have considered the introduction of cubic Lorentz-violating terms in 
the fermionic dispersion relation. We have concluded that the two possible Lorentz-violating 
parameters yield different terms in the fermionic dispersion relation, both 
cubic in the momentum operator components. In the non-relativistic limit, we have 
found the two leading order terms altering the equations of motion for fermions and  
determined the effect of these terms in the ${}^9\textrm Be^+$ ion's energy spectrum. 
Using the method developed 
by Weinberg and the experimental result of Bollinger \emph{et al.}, we have obtained new bounds on the 
value of the parameters $\eta_1$ and $\eta_2$ for neutrons. We have determined $|\eta_1|\lesssim6\times10^{-3}$ 
and $|\eta_2|\lesssim3$ for a time-like background Lorentz-violating four-vector, $|\eta_1|\lesssim2\times10^{-5}$ 
and $|\eta_2|\lesssim1\times10^{-8}$ for a space-like four-vector, and $|\eta_1|\lesssim2\times10^{-8}$ 
and $|\eta_2|\lesssim1\times10^{-8}$ for a light-like four-vector. 

The values of the Lorentz-violating parameters $\eta_1$ and $\eta_2$ are, hence, 
highly dependent on the form of the background four-vector, particularly on its 
spatial components. Bollinger \emph{et al.} experimental results are consistent with high values 
for these parameters, especially $|\eta_2|$, in the case where the spatial components of 
$n^{\mu}$ have small values in the laboratory frame, $n\sim10^{-3}$ (a time-like 
background four-vector). On the other hand, this experiment yields quite strong constraints when 
$n\sim1$ (a space-like or light-like background four-vector). 

In general, $n^{\mu}$ may have different spatial components in the laboratory frame due to the motion of the Earth with respect to the cosmic frame where the background four-vector has a simple form. If some of these components are further suppressed, the upper bounds on the values of the Lorentz-violating parameters will be larger than the ones presented above.

In any case, it is somewhat striking that 15 yr-old experiments like the one considered in 
this Letter can lead to relevant upper bounds for these parameters and shed some light
on the physics of very high energy scales.

\vskip 0.2cm

%%%%%%%%%%%%%%%%%%%%%%%%%%%%%%%%%%%%%%%%%%%%%%%%%%%%%%%%%%%%%%%%%%%%%%%%%%%%%%%%%%%%%%%%%%%%%%%%%%%%%%%%%%%%%%%%%%%%%%%%%%%%%%%%
\centerline{\bf {Acknowledgments}}

\vskip 0.2cm

\noindent The authors would like to thank David Mattingly for his useful comments and suggestions.

\vfill

%%%%%%%%%%%%%%%%%%%%%%%%%%%%%%%%%%%%%%%%%%%%%%%%%%%%%%%%%%%%%%%%%%%%%%%%%%%%%%%%%%%%%%%%%%%%%%%%%%%%%%%%%%%%%%%%%%%%%%%%%%%%%%%%
%\newpage

\end{document}